# AA stacking, tribological and electronic properties of double-layer graphene with krypton spacer

Andrey M. Popov[1,a)], Irina V. Lebedeva[2,b)], Andrey A. Knizhnik[3,4], Yurii E. Lozovik[1,5,c)], Boris V. Potapkin[3,4], Nikolai A. Poklonski[6,d)], Andrei. I. Siahlo[6], Sergey A. Vyrko[6]

[1]*Institute for Spectroscopy, Russian Academy of Sciences, Fizicheskaya Street 5, Troitsk, Moscow, 142190, Russia*

[2]*Nano-Bio Spectroscopy Group and ETSF Scientific Development Centre, Departamento de Física de Materiales, Universidad del País Vasco UPV/EHU, Avenida de Tolosa 72, E-20018 San Sebastián, Spain*

[3]*National Research Centre "Kurchatov Institute", Kurchatov Square 1, Moscow, 123182, Russia,*

[4]*Kintech Lab Ltd., Kurchatov Square 1, Moscow, 123182, Russia,*

[5]*Moscow Institute of Physics and Technology, Institutskii pereulok 9, Dolgoprudny, Moscow Region, 141700, Russia*

[6]*Physics Department, Belarusian State University, Nezavisimosti Ave. 4, 220030 Minsk, Belarus*

Structural, energetic and tribological characteristics of double-layer graphene with commensurate and incommensurate krypton spacers of nearly monolayer coverage are studied within the van der Waals-corrected density functional theory. It is shown that when the spacer is in the commensurate phase, the graphene layers have the AA stacking. For this phase, the barriers to relative in-plane translational and rotational motion and the shear mode frequency of the graphene layers are calculated. For the incommensurate phase, both of the barriers are found to be negligibly small. A

---

a) Electronic mail: popov-isan@mail.ru

b) Electronic mail: liv_ira@hotmail.com

c) Electronic mail: lozovik@isan.troitsk.ru

d) Electronic mail: poklonski@bsu.by

considerable change of tunneling conductance between the graphene layers separated by the commensurate krypton spacer at their relative subangstrom displacement is revealed by the use of the Bardeen method. The possibility of nanoelectromechanical systems based on the studied tribological and electronic properties of the considered heterostructures is discussed.

## I. INTRODUCTION

The discovery of graphene[1] sparked tremendous efforts in development of graphene-based nanometer-scale systems. The most important systems realized recently include AA-stacked bilayer[2] and multilayer[3] graphene and double-layer graphene,[4–10] i.e. the system consisting of two graphene layers separated by a dielectric spacer. While in bilayer graphene, the interlayer distance is about 3.4 Å and is close to that in graphite, the distance between the layers in double-layer graphene is determined by the thickness of the dielectric spacer.

Both AA-stacked bilayer graphene and double-layer graphene represent significant interest for studies of fundamental phenomena and practical applications. Electronic and optical properties of AA-stacked bilayer graphene were predicted to be very different from those of bilayer graphene with the ordinary AB Bernal stacking.[11–15] The quantum spin Hall effect,[11] spontaneous symmetry violations,[12] low-energy electronic spectra[13] and magneto-optical absorption spectra[14] for AA-stacked bilayer graphene and metal-insulator transition[15] for doped AA-stacked bilayer graphene were considered. A field-effect transistor consisting of two graphene layers with the nanometer-scale dielectric spacer between the layers was implemented.[4,5] A tunable metal-insulator transition was observed in double-layer graphene heterostructures.[6] Electron tunneling between graphene layers separated by an ultrathin boron nitride barrier was investigated.[7] Double-layer graphene heterostructures were also used to determine Fermi energy, Fermi velocity and Landau level broadering.[8] Measurements of Coulomb drag of massless fermions in double-layer graphene heterostructures were reported[9,10] and the theory of this phenomenon was considered.[16–22] Theoretical studies of electron-hole pairs condensation in a double-layer graphene were presented.[23–30]

Since the electronic properties of double-layer graphene with a thin dielectric spacer depend strongly on the stacking of the graphene layers, production of double-layer graphene with the controllable stacking is essential for its possible applications. Several types of double-layer graphene heterostructures with different dielectric spacers between the layers have been realized up to now. Namely, graphene layers can be separated by a few-nanometer $Al_2O_3$[8] or $SiO_2$[9] spacers, by one[5,7] or several[5–7,10] atomic boron nitride layers, and by a layer of adsorbed molecules.[4] However, most of these implementations do not allow to control the stacking of graphene layers. $Al_2O_3$[8] or $SiO_2$[9] spacers are not layered materials and, therefore, they do not make it possible to produce double-layer graphene not only with a given stacking of graphene layers but also with a given interlayer distance. As for the boron nitride spacer, the lattice constant of this material is 2% greater than the lattice constant of graphene.[31] The recent study of the commensurate-incommensurate phase transition in bilayer graphene with one stretched layer[32] revealed that even such a small mismatch of 2% between the lattice constants is sufficient for the transition to the incommensurate phase with a pattern of alternating commensurate and incommensurate regions so that the same relative position of the layers in the whole system is feasible only for a small size of the layers. In fact, the spatially inhomogeneous biaxial compressive strain has been observed lately in graphene/hexagonal boron nitride heterostructure.[33] As for the AA-stacked bilayer graphene without any spacer, observations of such a structure are restricted to the cases of bilayer graphene with a common folded edge[2] and AAAA-stacked regions of multilayer graphene on the C-terminated SiC substrate.[3] In the case of nearly coincident edges of a graphene flake and a graphene layer, stable and metastable positions of the flake differing both from the AA and AB stacking were found to be possible as a result of the trade-off between the edge-edge interaction and the van der Waals interaction.[34]

In the present paper, we suggest that a new type of graphene-based heterostructures, double-layer graphene with controllable stacking of the graphene layers, can be produced by the use of the layer of adsorbed atoms or molecules commensurate with the graphene layers as a spacer. It is well known that krypton can form commensurate layers on graphite (Ref. 35, 36 and references therein). Therefore, we consider this inert gas as a candidate for the commensurate spacer. The van der Waals-corrected density functional theory is applied to reveal the stacking of graphene layers separated by the commensurate krypton spacer and to calculate the barriers to relative motion of these layers and the shear mode frequencies. We show that the AA stacking of graphene layers can be realized in this heterostructure for an arbitrary substrate or in the suspended system and for an arbitrary size of neighbouring graphene layers, structure and relative position of their edges. Recently we have revealed frictionless tribilogical behavior for double-layer graphene with the incommensurate argon spacer.[37] However, our calculations have shown that the commensurate phase of the argon spacer between graphene layers is much less stable than the incommensurate phase and, therefore, should be difficult to obtain. The present paper is devoted to double-layer graphene with the commensurate spacer, i.e. to the heterostructure with radically different tribological and electronic properties.

The previous calculations demonstrated that the electronic structure of twisted bilayer graphene changes considerably with changing the twist angle.[38] The tunneling conductance between the layers of bilayer graphene changes by several times upon relative displacement of the layers[39,40] and by an order of magnitude upon relative rotation of the layers.[39] Here we use the Bardeen method to calculate 2D maps of tunneling conductance between graphene layers of bilayer graphene and double-layer graphene with the commensurate krypton spacer as a function of coordinates describing relative in-plane displacements of the layers. Possible applications of the revealed tribological and electronic properties of double-layer graphene with the commensurate and incommensurate krypton spacers in nanoelectromechanical systems (NEMS) are discussed.

The paper consists of the following parts. In Sec. II, we give the details of van der Waals-corrected density functional theory calculations. Sec. III is devoted to the analysis of structural and tribological properties of double-layer graphene with the krypton spacer. Sec. IV presents the results of calculations of tunneling conductance between krypton-separated graphene layers. In Sec. V, we consider the possibility of experimental realization and application of the studied heterostructure in NEMS and summarize our conclusions.

## II. COMPUTATIONAL DETAILS

Analysis of structural and tribological properties of krypton-separated double-layer graphene has been performed using the VASP code.[41] The performance of three approaches with the correction for the van der Waals interaction has been compared: (1) the DFT-D2 method[42] with the generalized gradient approximation (GGA) density functional of Perdew, Burke, and Ernzerhof[43] corrected with the dispersion term (PBE-D), (2) the vdW-DF method[44] with the optPBE-vdW exchange functional,[45,46] and (3) the vdW-DF2 method.[46,47] The basis set consists of plane waves with the maximum kinetic energy of 500-800 eV. The interaction of valence electrons with atomic cores is described using the projector augmented-wave method (PAW).[48] A second-order Methfessel–Paxton smearing[49] of the Fermi surface with a width of 0.1 eV is applied. The energy convergence tolerance for electronic self-consistent loops is $10^{-7}$ eV.

Two krypton spacers of different structure are considered. The spacer A is a krypton layer commensurate with the graphene layers (FIG. 1) and corresponds to the double-layer graphene with the krypton to carbon ratio Kr:C = 1:12. In this spacer, the distance between adjacent krypton atoms is 7% greater than the equilibrium distance in the isolated krypton layer of 4.00 Å for the DFT-D2 method, 5% greater than the equilibrium distance in the isolated krypton layer of 4.07 Å for the vdW-DF2 method and 4% greater than the equilibrium distance in the isolated krypton layer of 4.12 Å for the optPBE-vdW method. The spacer B has krypton atoms with inequivalent positions of krypton atoms on the graphene lattice within the model cell (FIG. 1) and corresponds to the krypton to carbon ratio Kr:C = 9:100. In this spacer, the distance between adjacent krypton atoms is only 3% greater than the equilibrium krypton-krypton distance according to the DFT-D2 method, only 1% greater than the equilibrium krypton-krypton distance for the vdW-DF2 method and very close to the equilibrium krypton-krypton distance for the optPBE-vdW method. The results below show that 9 krypton atoms with inequivalent positions for the spacer B are sufficient for the dramatic change of tribological characteristics of the system. Thus, we consider the spacer B as a prototype of the incommensurate spacer since we are not able to simulate incommensurate krypton-graphene heterostructures under the periodic boundary conditions (PBCs) directly.

An oblique-angled simulation cell is considered. The model cell has two equal sides at an angle of 60º and a perpendicular side of 20 Å length. The length of two equal sides is 7.402 Å for the spacer A and 12.34 Å for the spacer B. Integration over the Brillouin zone is performed using the Monkhorst-Pack method[50] with 12 x 12 x 1 – 24 x 24 x 1 and 7 x 7 x 1 - 14 x 14 x 1 k-point sampling for the spacers A and B, respectively. The structures of the graphene layers are separately geometrically optimized with the energy convergence tolerance of 0.001 meV/atom. After that the structures of the graphene layers are considered as rigid. Account of structure deformation induced by the interlayer interaction was shown to be inessential for the values of the barriers to relative motion of graphene-like layers, such as the interwall interaction of carbon nanotubes[51] and the intershell interaction of carbon nanoparticles.[52,53] Thermal fluctuations were also shown to have no influence on these barriers (by molecular dynamics simulations of orientational melting of double-shell carbon nanoparticles[52,53] and diffusion of a graphene flake on a graphene layer[54,55]). Furthermore, we do not consider here the possibility of symmetry distortions of the commensurate krypton spacer A (such a consideration would require the choice of PBCs compatible with each certain symmetry distortion) and treat it as rigid. The numerous studies of commensurate krypton layers on graphite (Ref. 35, 36 and references therein) and lateral surfaces of carbon nanotubes[56] have not revealed any symmetry distortions. The negligibly small barriers are found below for relative motion of graphene layers separated by the spacer B with the rigid structure. The distortions of the spacer structure can only decrease these barriers and thus do not affect this qualitative result.

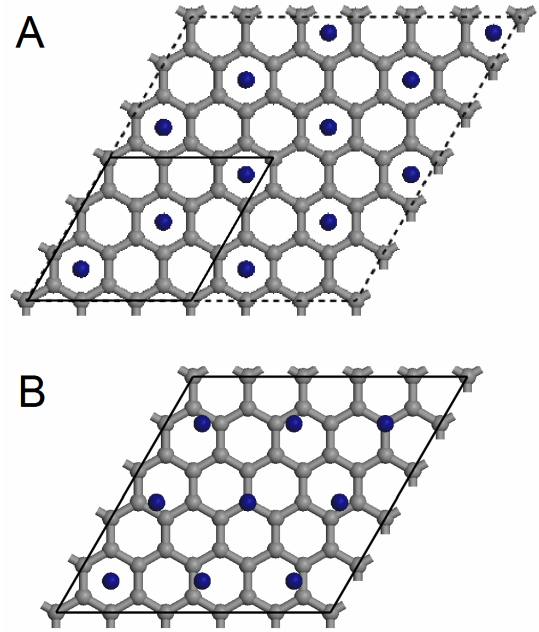

FIG. 1. Structures of krypton-separated double-layer graphene with the spacers A (Kr:C = 1:12) and B (Kr:C=9:100). Only one graphene layer is shown for clarity. Carbon atoms are coloured in gray, krypton atoms are coloured in dark blue. The model cells are indicated by solid lines.

The relative positions of the graphene layers and the krypton spacers A and B with the rigid structure corresponding to the potential energy minima are found through calculations of dependences of the potential energy on in-plane displacements of each of the graphene layers relative to the spacer and distances between the graphene layers and the spacer. The error in the total interaction energy due to the optimization of interlayer distances does not exceed 0.001 meV per carbon atom of one of the graphene layers. The contributions of interactions between the graphene layers, between each of the graphene layers and the krypton spacer, and inside the krypton spacer to the total interaction energy of the krypton-separated double-layer graphene (the interactions inside the graphene layers are excluded from this quantity) for found minimum energy positions are also considered. To evaluate these contributions the energies of the systems consisting of two graphene layers at the same interlayer distance and the in-plane relative position as in krypton-separated double-layer graphene but without the krypton spacer, the isolated graphene layer and the isolated krypton spacer are found.

The convergence on the number of k-points in the Brillouin zone and the maximum kinetic energy of plane waves was tested previously for bilayer graphene.[57] Additional convergence tests have been also performed for krypton-separated double-layer graphene. Increasing the number of k-points from 12 x 12 x 1 to 24 x 24 x 1 for the spacer A and from 7 x 7 x 1 to 14 x 14 x 1 for the spacer B and simultaneously increasing the maximum kinetic energy of plane waves from 500 to 800 eV leads to changes in the total interaction energy of double-layer graphene by less than 0.004 meV per carbon atom of one of the graphene layer. At the same time, the barrier to relative motion of the commensurate krypton spacer A and each of the graphene layers changes by less than 0.01 meV per carbon atom. Stretching or compressing the graphene layers by 0.5% from the ground state results in changes of the total interaction energy of krypton-separated graphene by less than 0.1 meV per carbon atom and changes of the barrier to relative motion of the commensurate krypton spacer A and each of the graphene layers by less than 0.01 meV per carbon atom. In Sec. III, we give the data obtained using the maximum kinetic energy of plane waves of 500 eV and the k-point grids of 12 x 12 x 1 and 7 x 7 x 1 for the spacers A and B, respectively.

## III. INTERACTION AND RELATIVE MOTION OF KRYPTON-SEPARATED GRAPHENE LAYERS

The structural characteristics and the total interaction energy of krypton-separated double-layer graphene as well as different contributions to this energy calculated using DFT-D2, optPBE-vdW and vdW-DF2 methods are listed in Table I. The data obtained for bilayer graphene using the same methods are also given for comparison.

Let us first discuss the results on structural properties of double-layer graphene. The equilibrium distances between the graphene layers separated by the spacers A and B calculated using all the methods considered lie in the range from 6.8 to 7.2 Å. These values are in qualitative agreement with the experimental value of ~6 Å for undetermined adsorbed molecules[4] and previous calculations for argon.[37] Nevertheless, the optPBE-vdW and vdW-DF2 methods predict slightly larger interlayer distances than the DFT-D2 method. To address the accuracy of the methods let us compare the results on structural properties of bilayer graphene (Table I) and of an isolated krypton layer with the available experimental data for graphite and krypton. The interlayer distance in graphite was measured to be 3.328 Å (Ref. 58) and 3.354 Å (Ref. 59). As seen from Table I, the optPBE-vdW and vdW-DF2 methods give the equilibrium interlayer distances for bilayer graphene greater than the experimental values by 0.1 – 0.2 Å, while the DFT-D2 method provides the interlayer distance smaller by 0.07 – 0.1 Å than these values. The experimental data on the equilibrium distance between neighbouring krypton atoms in few-layer and bulk krypton vary from 3.97 to 4.10 Å (Ref. 35, 36 and references therein). The equilibrium krypton-krypton distance of 4.00 Å calculated for the isolated krypton layer using the DFT-D2 method corresponds to the lower bound of this range, whereas the optPBE-vdW and vdW-DF2 methods give the distances of 4.12 Å and 4.07 Å, respectively, at the upper bound. Thus, the DFT-D2, optPBE-vdW and vdW-DF2 methods have comparable accuracy for description of structural properties of the van der Waals-bound systems considered here. The optPBE-vdW and vdW-DF2 methods tend to overestimate equilibrium distances, while the DFT-D2 method tends to underestimate equilibrium distances. Therefore, the interval of the equilibrium interlayer distances from 6.8 to 7.2 Å calculated for krypton-separated double-layer graphene using different methods should enclose the experimental data.

Table I. Calculated equilibrium distance $d_{gr-gr}$ between the graphene layers, total interaction energy $E_b$, interaction energy $E_{gr-Kr}$ between the krypton and graphene layers, interaction energy $E_{gr-gr}$ between the graphene layers, interaction energy $E_{Kr-Kr}$ between krypton atoms and barrier $\Delta E_{gr-gr}$ to relative motion of the graphene layers per carbon atom of one of the graphene layers for krypton-separated double-layer graphene. Calculated magnitude $\Delta E_{max}$ of corrugation of the potential energy relief and barrier $\Delta E_{gr-Kr}$ to translational motion of a single graphene layer relative to the krypton spacer per carbon atom at the graphene-krypton distance $d_{gr-Kr} = d_{gr-gr}/2$ are given. The data for bilayer graphene are also given for reference.

| Kr:C | $d_{gr-gr}$ (Å) | $E_b$ (meV) | $E_{gr-Kr}$ (meV) | $E_{gr-gr}$ (meV) | $E_{Kr-Kr}$ (meV) | $\Delta E_{max}$ (meV) | $\Delta E_{gr-Kr} = \Delta E_{gr-gr}$ (meV) |
|---|---|---|---|---|---|---|---|
| DFT-D2 | | | | | | | |
| 1:12 | 6.82 | -56.7 | -19.9 | -3.8 | -13.0 | 1.62 | 1.44 |
| 9:100 | 6.94 | -59.6 | -20.4 | -3.6 | -15.3 | <10$^{-5}$ | <10$^{-5}$ |
| bilayer graphene | 3.26 | -50.4 | | -50.4 | | 19.1 | 2.0 |
| optPBE-vdW | | | | | | | |
| 1:12 | 7.06 | -81.4 | -30.0 | -3.6 | -17.9 | 1.02 | 0.93 |
| 9:100 | 7.13 | -86.3 | -31.5 | -3.5 | -19.8 | | |
| bilayer graphene | 3.46 | -59.7 | | -59.7 | | 9.7 | 1.0 |
| vdW-DF2 | | | | | | | |
| 1:12 | 7.08 | -63.2 | -23.9 | -1.9 | -13.5 | 1.11 | 1.03 |
| 9:100 | 7.15 | -67.6 | -25.1 | -1.8 | -15.4 | | |
| bilayer graphene | 3.54 | -48.9 | | -48.9 | | 7.6 | 0.8 |

Let us now proceed with the discussion of relative stability of double-layer graphene with the spacers A and B and bilayer graphene. All the methods considered here agree that both for the spacers A and B, the total interaction energy of double-layer graphene is higher in magnitude than for bilayer graphene without any spacer (Table I). However, the quantitative data on the energies of these structures are rather different for different calculation methods. While according to the optPBE-vdW and vdW-DF2 methods, krypton-separated graphene with the spacers A and B is more stable than bilayer graphene by 14 – 27 meV per carbon atom of one of the graphene layers, for the DFT-D2 method, this energy difference is only 6 – 9 meV per carbon atom of one of the graphene layers. To consider the accuracy of the methods used we compare the calculated contributions to the total interaction energy of krypton-separated double-layer graphene and the calculated interlayer binding energy in bilayer graphene with the experimental data.

The experimentally measured adsorption energies of krypton atoms on graphite lie in the interval from -117 meV to -126 meV per krypton atom (Ref. 60-62 and references therein). For the krypton-graphene interaction in double-layer graphene, the optPBE-vdW and vdW-DF2 methods give the energies of -180 meV and -144 meV per krypton atom and one graphene layer for the spacer A and of -175 meV and -140 meV per krypton atom for the spacer B, respectively, i.e. these methods strongly overestimate the magnitude of krypton-graphene interaction. On the other hand, the DFT-D2 method provides the krypton-graphene energies of -120 meV and -113 meV per krypton atom and one graphene layer for the spacers A and B, respectively. Thus, the DFT-D2 method is more accurate in the description of krypton-graphene interaction than the optPBE-vdW and vdW-DF2 methods.

The krypton-krypton interaction energy in the krypton commensurate layer with the same krypton coverage as in the spacer A can be deduced from the the exponential factor in the temperature dependence of the critical pressure for the commensurate-incommensurate phase transition to be around -50 meV per krypton atom.[63] The optPBE-vdW method gives the krypton-krypton interaction energy of -107 meV per krypton atom for the spacer A and of -110 meV per krypton atom for the spacer B, i.e. much greater in magnitude than the values estimated from the experimental data. According to the DFT-D2 and vdW-DF2 methods, the krypton-krypton interaction energy is -78 meV and -81 meV per krypton atom for the spacer A and -84 meV and -86 meV per krypton atom for the spacer B, respectively. So these methods also overestimate the krypton-krypton interaction energy but the discrepancy with the experimental data is smaller.

Let us finally compare the accuracy of the methods considered with respect to the interlayer interaction in bilayer graphene. The latest experimental measurements for graphite[64] gave the interlayer binding energy of -52±5 meV per carbon atom. It is seen from Table I that the optPBE-vdW strongly overestimates the interlayer binding energy in bilayer graphene, while the DFT-D2 and vdW-DF2 methods provide reasonable values of this energy. From all these considerations, the DFT-D2 approach seems to be the most reliable for the analysis of relative energies of graphene-based heterostructures among the methods used here.

Let us now address the relative stability of the commensurate and incommensurate phases of the krypton spacer. All the methods considered predict that the total interaction energy is higher in magnitude for the spacer B with the higher krypton coverage than for the spacer A (Table I). According to the DFT-D2 calculations, the incommensurate spacer B is preferred in energy compared to the spacer A only by 3 meV per carbon atom of one of the graphene layers. Extrapolating the krypton-graphene and graphene-graphene binding energies to the limit of krypton coverage corresponding to the equilibrium krypton-krypton distance in the isolated krypton layer of 4.0 Å, we can estimate that the incommensurate phase should be preferred over the commensurate one by no more than 4 meV per carbon atom of one of the graphene layers. This energy difference between the commensurate and incommensurate spacers is more than 3 times less than in the case of argon between graphene layers, for which the incommensurate phase is much more favorable than the commensurate one.[37] The similarly small energy difference between the heterostructures with the krypton spacers A and B of 4.4 meV per carbon atom of one of the graphene layers follows from the relatively accurate vdW-DF2 method (Table I). Therefore, it should be much easier to obtain the commensurate phase of krypton in the double-layer graphene than the commensurate phase of argon. This conclusion is in agreement with the experimental observations of argon and krypton adsorption on graphite and carbon nanotubes. While the commensurate phase of krypton is found in relatively wide ranges of ambient pressures and temperatures (Refs. 35, 36, 56 and references therein), the commensurate phase of argon on graphite[65–67] or carbon nanotubes[56] has not been detected so far. The possibility of decay of double-layer graphene with the commensurate krypton spacer into regions of double-layer graphene with the incommensurate krypton layer and bilayer graphene is considered in Sec. V.

To obtain the barrier to relative motion of the graphene layers in double-layer graphene with the krypton spacer the potential energy reliefs, i.e. the dependences of interaction energy between the graphene layer and the krypton spacer on coordinates describing their relative in-plane displacements, have been calculated for the krypton-graphene distance that is equilibrium for the double-layer graphene. The potential energy relief for the commensurate krypton spacer A calculated using the DFT-D2 approach is shown in Figure 2. For this spacer, the minima of this potential relief correspond to the relative positions of the spacer and graphene layers in which krypton atoms are placed in the centers of hexagons of the graphene lattice. For the krypton spacer B with nonequivalent positions of krypton atoms within the model cell, the magnitude of corrugation of the potential energy relief is found to be below $10^{-5}$ meV per carbon atom (similar to incommensurate argon spacers[37]) and is too small to determine the relative position of the spacer and graphene layers corresponding to the energy minimum.

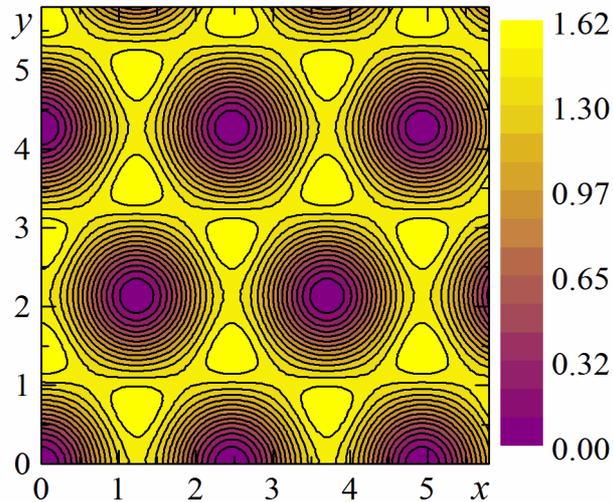

FIG. 2. Interaction energy (in meV) between the graphene layer and the commensurate krypton spacer A (Kr:C=1:12) per carbon atom as a function of the relative position of the krypton layer $x$ and $y$ (in Å; $x$ and $y$ axes are chosen along the zigzag and armchair directions, respectively) calculated using the DFT-D2 method. The position $x = 0$ and $y = 0$ corresponds to the ground state of the commensurate krypton layer on the graphene layer. The equipotential lines are drawn with a step of 0.108 meV.

The magnitudes of corrugation of the potential energy relief and the barriers to translational motion of each of the graphene layers relative to the krypton layer are given in Table I. These data show considerable scatter for different methods. As it was shown in our previous publication, though the correction for the van der Waals interaction does not contribute much to the roughness of the potential energy relief, this relief is very sensitive to the interlayer distance.[57] Due to the differences in the interlayer distances, the calculated barriers vary by 40% from the DFT-D2 method to the optPBE-vdW method. The same as for interlayer distances, the barriers calculated using different methods mark the interval that should enclose the experimental data. To address the accuracy of different methods with respect to the description of tribological properties of layered graphene-based structures let us compare the results for bilayer graphene. On the basis of experimentally measured shear mode frequencies for few-layer graphene and graphite, the barrier to relative motion of graphene layers and the magnitude of corrugation of the potential energy relief for bilayer graphene were estimated to be 1.7 meV and 15 meV per carbon atom of one of the graphene layers, respectively.[68] The optPBE-vdW and

vdW-DF2 methods underestimate these quantities by 40% and 50%, respectively, whereas the DFT-D2 method overestimates these quantities by only 25% (Table I). Thus, it can be expected that the DFT-D2 method is more accurate in the prediction of tribological properties of krypton-separated double-layer graphene.

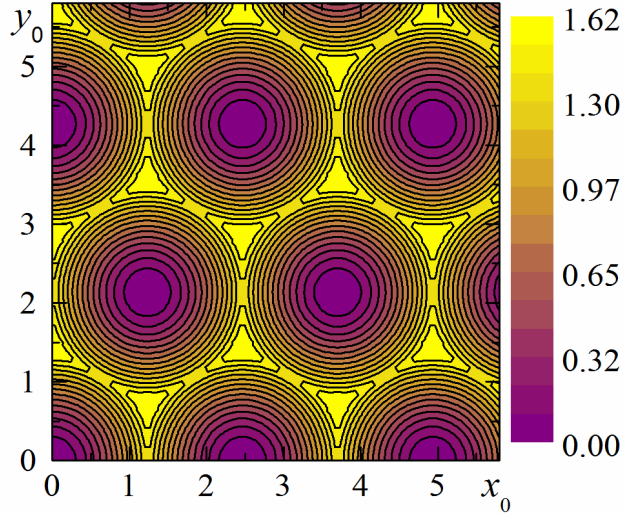

FIG. 3. Interaction energy (in meV) of graphene layers in double-layer graphene with the commensurate krypton spacer A (Kr:C=1:12) per atom of one of the graphene layers as a function of the relative position of the layers $x_0$ and $y_0$ (in Å; $x$ and $y$ axes are chosen along the zigzag and armchair directions, respectively) obtained using Equation 2 parameterized on the basis of the DFT-D2 calculations. The position $x_0 = 0$ and $y_0 = 0$ corresponds to the AA stacking of the graphene layers. The equipotential lines are drawn with a step of 0.108 meV.

The same as for commensurate argon spacers,[37] the shape of the potential energy relief for the commensurate spacer A is well described by the expression containing only the first Fourier components (FIG. 2)

$$U_{gr-Kr}(x, y, z) = U_1(z)\big(3 - 2\cos(k_1 x)\cos(k_2 y) - \cos(2k_2 y)\big) + U_0(z), \quad (1)$$

where $k_1 = 2\pi / a_0$, $k_2 = 2\pi / (\sqrt{3} a_0)$, $a_0 = 2.46$ Å is the lattice constant of graphene, $x$ and $y$ are coordinates corresponding to the in-plane relative displacement of the krypton spacer and graphene layer ($x$ and $y$ axes are chosen along the zigzag and armchair directions, respectively) and $z$ is the distance between the krypton spacer and the graphene layer. From the DFT-D2 calculations, the energy parameters $U_0$ and $U_1$ are found to be $U_0 = -20.0$ meV and $U_1 = 0.36$ meV per carbon atom. The relative root-mean square deviation $\delta U / U_1$ of the potential energy relief obtained using Equation 1 from the one obtained by the DFT-D2 calculations is only 1%. From this equation, the magnitude of corrugation of the potential energy relief and the barrier to relative motion of the graphene and krypton layers can be found as $\Delta E_{max} \approx 4.5 U_1 = 1.62$ meV and $\Delta E_{\text{gr-Kr}} \approx 4U_1 = 1.44$ meV per carbon atom, respectively. It should be also noted that the first Fourier components were previously shown to be sufficient for the description of the potential reliefs of interlayer interaction energy in bilayer graphene[54,55,57,68,69] and interwall interaction energy in carbon nanotubes.[51,70,71]

As shown previously for argon spacers,[37] the contribution of graphene-graphene interaction to corrugations of the potential energy relief in double-layer graphene does not exceed 0.003 meV per carbon atom of one of the layers. This is more than two orders of magnitude smaller than the magnitude of corrugation obtained for the interaction between the graphene layers and the krypton spacer A (Table 1). Therefore, the variation of the total interaction energy upon relative motion of the graphene layers can be obtained as the sum of variations of interaction energies between the krypton layer and each of the graphene layers, both of which can be approximated by Equation 1. The minimal energy of the system with the relative position of graphene layers $x_0, y_0, z_0$ can, therefore, be calculated as

$$\delta U_{\text{gr-gr}}(x_0, y_0, z_0) = \min_{x,y} \left[ U_{\text{gr-Kr}}(x, y, z_0/2) + U_{\text{gr-Kr}}(x_0 - x, y_0 - y, z_0/2) - 2U_0(z_0/2) \right] \quad (2)$$

The potential energy relief calculated using Equation 2 is given in FIG. 3. We should emphasize that as opposed to bilayer graphene with the AB stacking of layers, the AA stacking of graphene layers is found here for the ground state of the double-layer graphene with the commensurate krypton spacer A.

In this case of relative displacement along the energy favourable zigzag direction (see FIG. 2 and FIG. 3), the expression (2) reduces to

$$\delta U_{\text{gr-gr}}(x_0, y_0, z_0) = 2U_1(z_0/2) \min_x \left[ 2 - \left( \cos(k_1 x) + \cos(k_1(x_0 - x)) \right) \right]$$
$$= 4U_1(z_0/2) \left[ 1 - \cos(k_1(x_0 - Na_0)/2) \right], \quad x_0 \in \left[ (N - 1/2)a_0; (N + 1/2)a_0 \right], \quad (3)$$

where $N$ is an integer. From this formula, it is seen that the barrier to relative motion of the graphene layers in double-layer graphene with the commensurate krypton layer is exactly equal to the barrier to relative motion of the krypton layer on one of the graphene layers and in the case of parametrization on the basis of the DFT-D2 calculations is $\Delta E_{\text{gr-gr}} = \Delta E_{\text{gr-Kr}} \approx 4U_1 = 1.44$ meV per carbon atom of one of the graphene layers (Table I).

For double-layer graphene with the incommensurate krypton spacer, the potential energy relief for relative motion of the graphene layers is extremely smooth due to the absence of barriers to relative motion of the incommensurate krypton spacer and each of the graphene layers. Therefore, the static friction force for relative motion of the graphene layers is negligibly small.

The expression (3) can be also used to estimate the shear mode frequency of graphene with $n$ layers separated by the commensurate krypton spacers corresponding to in-plane vibrations of adjacent graphene layers in opposite directions. Assuming that the krypton atoms do not move in this mode, its frequency can be found as

$$f = \frac{1}{2\pi}\sqrt{\frac{n-1}{\mu}\frac{\partial^2 \delta U_{\text{gr-gr}}}{\partial x_0^2}} = \frac{1}{a_0}\sqrt{\frac{(n-1)U_1}{\mu}}, \tag{4}$$

where $\mu = mp/2$ for $n = 2p$ and $\mu = mp(p+1)/(2p+1)$ for $n = 2p+1$, $m$ is the mass of a carbon atom and $p$ is an integer. This formula looks just the same as for graphene bilayer.[57,68] However, here $U_1$ characterizes the krypton-graphene interaction instead of the graphene-graphene interaction. For double-layer graphene with the commensurate krypton spacer A, the formula parameterized on the basis of the DFT-D2 calculations gives the frequency $f = 10.5\,\text{cm}^{-1}$, which is three times smaller than for the graphene bilayer.[57, 68,69,72] This can be explained by the one-order difference in the magnitudes of corrugation of the potential energy reliefs for krypton-separated double-layer graphene and graphene bilayer.[57, 68,69]

For the material consisting of alternating graphene layers and commensurate krypron spacers A, the formula (4) parameterized on the basis of the DFT-D2 calculations gives the shear mode frequency $f = 14.8\,\text{cm}^{-1}$, three times smaller than for graphite.[68] The shear modulus of this material can be estimated as

$$C_{44} = \frac{4d_{\text{gr-gr}}}{\sqrt{3}a_0^2}\frac{\partial^2 \delta U_{\text{gr-gr}}}{\partial x_0^2} = \frac{4\pi^2 d_{\text{gr-gr}} mf^2}{\sqrt{3}a_0^2} \approx 0.98\,\text{GPa} \tag{5}$$

and is four-five times smaller than the experimental values for the shear modulus of graphite (see Ref. 72 and references therein).

The calculated potential energy relief for the commensurate krypton spacer also allows to get the upper bound estimate for the barrier to relative rotation of graphene layers to incommensurate states. Supposing that the argon layer stays rigid during the rotation and takes equivalent positions with respect to each of the graphene layers (at equal angles), the barrier to relative rotation of graphene layers in the double-layer graphene should be twice greater than the barrier to rotation of each of the graphene layer relative to the rigid argon layer. The latter barrier can be found as the difference between the average of expression (1) over all in-plane displacements of the krypton layer $x$, $y$ and its minimum value for the equilibrium krypton-graphene distance $\langle U_{\text{gr-Kr}}\rangle_{x,y} - U_0 = 3U_1$. In this way on the basis of the DFT-D2 calculations, we get the value $\Delta E_{\text{gr-gr}}^{\text{rot}} \approx 6U_1 = 2.2$ meV per carbon atom of one of the graphene layers for the barrier to relative rotation of the graphene layers in double-layer graphene. However, in reality, such a rotation would be accompanied by shrinkage of the krypton layer and release of the additional energy of a few meV per carbon atom. Thus, this value $\Delta E_{\text{gr-gr}}^{\text{rot}}$ corresponds to the upper bound estimate of the barrier for rotation.

## IV. TUNNELING CONDUCTANCE BETWEEN KRYPTON-SEPARATED GRAPHENE LAYERS

As well known, the tunneling conductance $G$ is proportional to the sum of squares of the amplitudes of tunneling matrix elements for all electron states at both sides of the tunneling transition.[73] To calculate tunneling matrix elements between states of the bottom ($\Psi_{\text{bot}}$) and top ($\Psi_{\text{top}}$) layers of bilayer graphene and between states of the bottom and top layers of double-layer graphene with the commensurate krypton spacer we have applied the Bardeen formalism described in detail in Ref. 40. According this formalism,[40,74] the matrix element between two states has the form

$$M_{\text{bot,top}}^{\mathbf{k}_{\text{bot}}=\mathbf{k}_{\text{top}}=\mathbf{K}} = \frac{\hbar^2}{2m_0} \int_S \left( \Psi_{\text{bot}}^* \nabla \Psi_{\text{top}} - \Psi_{\text{top}}^* \nabla \Psi_{\text{bot}} \right) d\mathbf{S}, \tag{6}$$

where $S$ is the overlap area between the graphene layers, $m_0$ is the electron mass in vacuum, $\hbar$ is the Planck constant. Two-dimensional wave vectors $\mathbf{k} = \mathbf{k}_{\text{bot}} = \mathbf{k}_{\text{top}} = \mathbf{K} = \left( \pm 2\pi/3a_0; 2\pi/\sqrt{3}a_0 \right)$ near the corners of the Brillouin zone are considered.

In the tight-binding approximation, the wave function of the bottom/top graphene layer takes the form[75]

$$\Psi_{\text{bot(top)}}(\mathbf{k},\mathbf{r}) = \frac{1}{\sqrt{N_G}} \sum_{g=1}^{N_G} \exp\left(i\mathbf{k}\mathbf{R}_g^{\text{bot(top)}}\right) \frac{1}{\sqrt{2}} \left( \chi\left(\mathbf{r} - \mathbf{R}_g^{\text{bot(top)}}\right) \pm \frac{\omega(\mathbf{k})}{|\omega(\mathbf{k})|} \chi\left(\mathbf{r} - \mathbf{R}_g^{\text{bot(top)}} - \mathbf{d}\right) \right). \tag{7}$$

Here $N_G$ is the number of elementary unit cells of graphene, $\mathbf{d} = (\mathbf{a}_1 + \mathbf{a}_2)/3 = \left(0; a_0/\sqrt{3}\right)$ is the vector between two inequivalent carbon atoms ($A$ and $B$), $\mathbf{a}_1 = \left(a_0/2; a_0\sqrt{3}/2\right)$ and $\mathbf{a}_2 = \left(-a_0/2; a_0\sqrt{3}/2\right)$ are the ground vectors of the graphene lattice (the signs + and − correspond to $\pi$ (bonding) and $\pi^*$ (antibonding) orbitals in graphene), $\mathbf{R}_g^{\text{bot(top)}} = g_x \mathbf{a}_1 + g_y \mathbf{a}_2$ is the radius vector of the $g$-th unit cell of the bottom/top graphene layer, $\mathbf{r}$ is the radius vector, $\omega(\mathbf{k}) = 1 + exp(i\mathbf{ka}_1) + exp(-i\mathbf{ka}_1)$, $\omega(\mathbf{k})/|\omega(\mathbf{k})| \approx 1$ near $K$-points of the Brillouin zone, $\chi(\mathbf{r})$ is the Slater $2p_x$-orbital of a carbon atom

$$\chi(\mathbf{r}) = \left(\frac{\xi^5}{\pi}\right)^{1/2} z \, exp\left(-\xi\sqrt{x^2 + y^2 + z^2}\right), \qquad (8)$$

where $\xi = 1.5679/a_B$ (Ref. 76), $a_B$ = 0.0529 nm is the Bohr radius, the $z$ axis is perpendicular to the $xy$ plane of graphene, $r = \sqrt{x^2 + y^2 + z^2}$ is the magnitude of the radius-vector $\mathbf{r}$ from the carbon atom center.

According to Ref. 40, the expression for $M_{bot,top}^{\mathbf{k}}$ can be written as

$$M_{bot,top}^{\mathbf{k}} = \frac{1}{2} \sum_{g=1}^{N_G^{bot}} exp(i\mathbf{k}\Delta\mathbf{R}_g)\left(\gamma_{A-A'_g} + \gamma_{B-A'_g} + \gamma_{A-B'_g} + \gamma_{B-B'_g}\right) = M_{bot,top}, \qquad (9)$$

where $N_G^{bot}$ is the number of unit cells of the top layer located at the distance less than $\Delta R_{max}$ from the considered unit cell of the bottom layer parallel to the graphene plane, $\gamma_{A(B)-A'_g(B'_g)}$ are the hopping integrals between atoms $A$ ($B$) of the considered unit cell of the bottom layer and $A'_g$ ($B'_g$) of the $g$-th unit cell of the top layer

$$\gamma_{A(B)-A'_g(B'_g)} = \frac{\hbar^2}{2m_0} \int_S \left(\chi_{bot}\frac{d\chi_{top}}{dz} - \chi_{top}\frac{d\chi_{bot}}{dz}\right) dS . \qquad (10)$$

Here $dS = dxdy$, $\chi_{top} = \chi\left(x - X_{A(B)}, y - Y_{A(B)}, d_{gr-gr}/2\right)$, $\chi_{bot} = \chi\left(x - \left(X_{A'_g(B'_g)} - x_0\right), y - \left(Y_{A'_g(B'_g)} - y_0\right), -d_{gr-gr}/2\right)$, $x$ and $y$ are the distances along the $x$ and $y$ axes between the considered atom of the bottom layer and the atom of the top layer, $x_0$ and $y_0$ are the displacements of the graphene layers along the $x$ and $y$ axes, $X_{A(B)}$ and $Y_{A(B)}$ are the coordinates of the $A$ and $B$ atoms in the elementary unit cell of the bottom layer of graphene ($X_A = X_B = 0$, $Y_A = 0$, $Y_B = a_0/\sqrt{3}$), $X_{A'_g(B'_g)}$ and $Y_{A'_g(B'_g)}$ are the coordinates of the $g$-th elementary unit cell of the top graphene layer, i.e. $X_{A'_g(B'_g)} = X_{A(B)} + g_x a_0$, $Y_{A'_g(B'_g)} = Y_{A(B)} - a_0/\sqrt{3} + g_y a_0 \sqrt{3}$ for the bilayer graphene and $X_{A'_g(B'_g)} = X_{A(B)} + g_x a_0$, $Y_{A'_g(B'_g)} = Y_{A(B)} + g_y a_0 \sqrt{3}$ for two graphene layers separated by krypton atoms. In calculations of matrix elements according to Eqs. (6) - (10), the interactions between elementary cells at distances less than $5a_0$ are taken into account.

FIG. 4 shows the calculated matrix elements as functions of the relative displacement of graphene layers for bilayer graphene (FIG. 4a) and krypton-separated double-layer graphene (FIG. 4b). In the ground state of bilayer graphene (Bernal structure, $x_0 = 0$ and $y_0 = 0$), the matrix element is found to be 0.136 eV, while for the AA-stacking of bilayer graphene ($x_0 = 0$ and $y_0 = -a_0/\sqrt{3}$), the matrix element is shown to be maximal and equal to 0.272 eV. In the ground state of krypton-separated double-layer graphene (AA-stacking), the matrix element is found to be only $2.79 \cdot 10^{-5}$ eV.

The function $M_{bot,top}(x_0, y_0)$ can be approximated as

$$M_{bot,top}(x_0, y_0) = M_0 + M_1 \cos\left(\frac{2\pi(x_0 - X_{A_0'})}{a_0}\right) \cos\left(\frac{2\pi(y_0 - Y_{A_0'})}{\sqrt{3}a_0}\right), \quad (11)$$

where $X_{A_0'} = 0$, $Y_{A_0'} = -a_0/\sqrt{3}$, $M_0 = 0.183$ eV, $M_1 = 0.090$ eV for bilayer graphene, and $X_{A_0'} = 0$, $Y_{A_0'} = 0$, $M_0 = 1.86 \cdot 10^{-5}$ eV and $M_1 = 9.32 \cdot 10^{-6}$ eV for krypton-separated double-layer graphene. The relative root-mean square deviations of the matrix elements obtained using Eq. (11) from the values calculated according to Eqs. (6) – (10) for bilayer graphene and for krypton-separated double-layer graphene are only 3.4% and 3.7%, respectively, while the maximal deviations are 10.7% and 13.5%, respectively.

The ratio of tunneling conductances between graphene layers with different stacking and interlayer distance equals the squared ratio of the corresponding matrix elements determined by Eq. (8). Thus, the tunneling conductance between the graphene layers of double-layer graphene with the commensurate krypton spacer in the ground state (the AA stacking of graphene layers) is about seven orders of magnitude smaller than the tunneling conductance between the layers of bilayer graphene in the ground state (the AB stacking). The dependences of the ratio of the tunneling conductance $G$ to the tunneling conductance $G_0$ of bilayer graphene in the ground state ($x_0 = 0$ and $y_0 = 0$) on the relative displacement of the layers for bilayer graphene and double-layer graphene with the commensurate krypton spacer A are shown in FIG. 4c and d, respectively. It is seen that the tunneling conductance between the graphene layers strongly depends on their relative position at the subnanometer scale, similar to the results obtained previously for bilayer graphene[40] and for double-walled carbon nanotubes.[77–79] In the both considered systems, the tunneling conductance reaches its maximum for the AA stacking, in which atoms of the graphene layers are located at the smallest distances to each other, while the minima of the tunneling conductance correspond to the SP stacking. However, the difference in the ground state stacking leads also to qualitatively different changes of the tunneling conductance at subangstrom in-plane relative displacements of the graphene layers from the ground state. Namely, for bilayer graphene with the AB stacking at the ground state, a decrease or an increase of the tunneling conductance is possible depending on the direction of the displacement, whereas for double-layer graphene with the commensurate krypton spacer, the AA stacking at the ground state corresponds to the maximal tunneling conductance and any in-plane relative displacement of the graphene layers causes a decrease of the conductance. Possible NEMS that can be based on this qualitatively different behavior of the tunneling conductance at the ground state are discussed below.

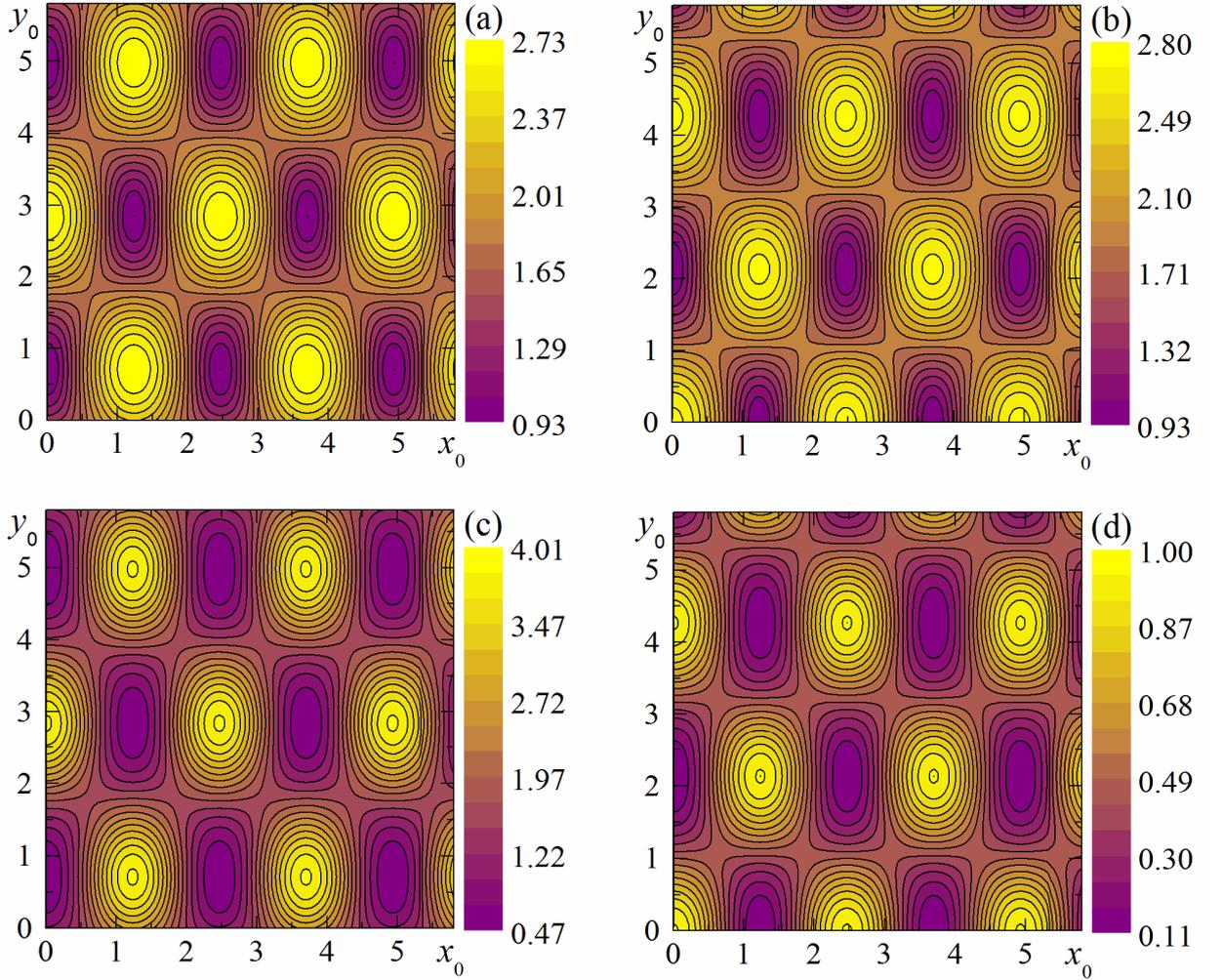

FIG. 4. Calculated (a, b) matrix element $M_{bot,top}$ (in (a) $10^{-1}$ eV and (b) $10^{-5}$ eV) and (c, d) relative tunneling conductance $G/G_0$ as functions of the relative displacement of graphene layers $x_0$ and $y_0$ (in Å; $x$ and $y$ axes are chosen along the zigzag and armchair directions, respectively) (a, c) in bilayer graphene and (b, d) in double-layer graphene with the commensurate krypton spacer A. The position $x_0 = 0$ and $y_0 = 0$ corresponds to the ground state of the considered systems: (a, c) the AB stacking of graphene layers in bilayer graphene and (b, d) the AA stacking of graphene

layers in krypton-separated double-layer graphene. The lines are drawn with a step of (a) 0.012 eV, (b) $1.3 \cdot 10^{-6}$ eV, (c) 0.25 and (d) 0.063.

## V. DISCUSSION AND CONCLUSION

We have considered the possibility to produce a new type of graphene-based heterostructure, double-layer graphene with controllable interlayer distance and stacking of the graphene layers, by using a layer of adsorbed inert atoms as a spacer. For this purpose, the density functional theory calculations of structural, energetic and tribological characteristics of the heterostructure consisting of two graphene layers separated by the commensurate (A) and incommensurate (B) krypton spacers have been carried out. The performance of the DFT-D2 method,[42] vdW-DF method[44] with the optPBE-vdW exchange functional[45,46] and vdW-DF2 method[47] has been compared. Though all the methods predict qualitatively the same results, the DFT-D2 method is shown to be more reliable in the quantitative description of energetic and tribological properties of double-layer and bilayer graphene.

The contributions of interaction between the krypton atoms, between the krypton spacer and the graphene layers, and between the graphene layers into the total interaction energy of the system have been obtained for both of the spacers. All the methods used agree that both of the considered structures of double-layer graphene are more stable than bilayer graphene, i.e. the escape of krypton atoms from these structures is not energetically favorable.

Furthermore, it is revealed that the structures with the commensurate and incommensurate krypton spacers are close in energy. Though the calculations show that the decay of double-layer graphene with the commensurate krypton spacer into regions of double-layer graphene with the incommensurate krypton spacer and bilayer graphene is slightly favourable energetically, this process clearly has a considerable barrier as the energy release occurs only after the graphene layers stick together. Let us estimate the activation energy for nucleation of the critical island of bilayer graphene in double-layer graphene with the commensurate spacer A. There are two contributions to this activation energy: (1) the excessive elastic energy of the curved graphene layers and (2) the energy required to compress the krypton layer in order to form the area of double-layer graphene free of krypton atoms. We denote the diameter of the critical island of bilayer graphene as $L$ and assume that to stick together both of the graphene layers get curved and the distance between them decreases by $2l \approx 7.0 - 3.4 = 3.6$ Å. The characteristic curvature radius of graphene layers in this area can be estimated as $R \approx (L/4)^2 / (l/2) = L^2 / (8l)$. Then the total elastic energy of the graphene is given by $E_{gr} \approx 2CL^2 / (\sigma R^2) = 128Cl^2 / (\sigma L^2)$ (see, for example, Ref. 80 and references therein), where $\sigma = 2.6$ Å$^2$ is the area per carbon atom in graphene and the coefficient $C$ was calculated for carbon nanotubes to be $C \approx 2$ eV·Å$^2$ per carbon atom.[80] The second contribution to the activation energy for nucleation of the critical island of bilayer graphene is related to the fact that to form this island it is necessary to free this area from krypton atoms and, correspondingly, to compress the krypton layer. The krypton coverage in the spacer A (the ratio of krypton atoms to carbon atoms in one of the graphene layers) is $\theta_{com} = 0.167$. Based on the DFT-D2 calculations, the krypton coverage in the incommensurate spacer with the equilibrium krypton-krypton distance is estimated to be $\theta_{in} = 0.189$. According to the same method, the energy required to compress the spacer A to the incommensurate structure with the equilibrium krypton-krypton distance without sticking the graphene layers is $\delta\varepsilon \approx 3$ meV per carbon atom of one of the graphene. Thus, the energy associated with compression of the krypton spacer without sticking the graphene layers can be estimated as $E_{Kr} \approx L^2 \theta_{in} \delta\varepsilon / (\sigma(\theta_{in} - \theta_A))$. Minimizing the sum of these two energies, we obtain an estimate of the characteristic size of the critical island of bilayer graphene in double-layer graphene with the spacer A

$$L_c \approx \left( \frac{128 C l^2 (\theta_{in} - \theta_{com})}{\delta\varepsilon \theta_{in}} \right)^{1/4} = 13 \text{ Å}. \tag{12}$$

For this size of the critical island, the activation energy can be estimated as $E_a = E_{Kr} + E_{gr} \approx 2 L_c^2 \theta_{in} \delta\varepsilon / (\sigma(\theta_{in} - \theta_{com})) \approx 3$ eV. This means that at room temperature double-layer graphene with the commensurate spacer should be stable for a very long time. Thus, we propose that both of the heterostructures of double-layer graphene with the commensurate and incommensurate krypton spacers can be obtained experimentally. To implement them it is sufficient to combine well established procedures of krypton deposition on graphene (analogously to the commensurate and incommensurate krypton layers on graphite[35,36] and carbon nanotubes[56]) and of transfer of graphene layers (see, for example, Ref. 5).

For double-layer graphene with the commensurate krypton spacer, considerable corrugations of the potential energy relief describing the relative in-plane displacements of the graphene layers are revealed. On the basis of the DFT-D2 calculations, the barrier for relative motion of the graphene layers in this heterostructure is found to be 1.44 meV per carbon atom, which is about 70% of the corresponding barrier in bilayer graphene.[57] Simultaneously, the changes in the tunneling conductance between the graphene layers at their relative displacement through this barrier are calculated here to be up to 90%. A set of NEMS based on the interaction and subangstrom relative motion of layers of bilayer graphene was proposed, including the nanoresonator[69], two different schemes of the force sensor[40,81] and the floating gate memory cell.[81] NEMS with analogous or other operational principles based on krypton-separated graphene layers can be elaborated. The Q-factor for subangstrom relative vibrations of the layers of bilayer graphene is rather small $Q = 30 - 150$.[69] The presence of the krypton spacer should lead to additional channels of energy dissipation of the mechanical oscillations arising after switching or measurement events in NEMS based on krypton-separated graphene layers. Thus, the considered heterostructure is perspective for elaboration of fast-acting sensors and other NEMS.

For double-layer graphene with the incommensurate krypton spacer, the potential energy relief describing the relative in-plane displacements of the graphene layers is shown to be extremely smooth and, therefore, the static friction force for relative motion of the graphene layers is negligibly small. Thus, double-layer graphene with the incommensurate krypton spacer is also suitable for NEMS based on free relative translational or rotational motion of the graphene layers proposed recently for argon-separated double-layer graphene.[37] In particular, the revealed static-friction-free relative in-plane motion of the graphene layers separated by the incommensurate krypton spacer allows us to propose that such a heterostructure can be perspective for elaboration of variable capacitors with the capacitance proportional to the overlap area of the layers. The tunneling conductance between the krypton-separated graphene layers is found to be seven orders of magnitude smaller than the tunneling conductance between the layers of bilayer graphene. The distance between the graphene layers separated by the incommensurate argon spacer consisting of two atomic layers was calculated to be 9.97 Å.[37] For this distance between the graphene layers, the tunneling conductance between the layers can be estimated to be 18 orders of magnitude less than for bilayer graphene. Thus, the leakage current of capacitors based on double-layer graphene with the krypton and argon spacers is sufficiently small to apply them in fast-acting nanodevices.

Up to now the AA stacking of graphene layers is found only for bilayer graphene with common folded edge[2] or local regions of multi-layer graphene on the C-terminated SiC substrate.[3] We have found that at the ground state of double-layer graphene with the commensurate krypton spacer, the AA stacking of graphene layers takes place. The AA-stacking can be realized for this heterostructure for an arbitrary size of neighbor graphene layers, structure and relative positions of their edges. A set of various phenomena was predicted for the AA-stacked bilayer graphene.[11–15] We believe that the considered AA-stacked double-layer graphene with the dielectric spacer holds great promise both for studies of fundamental phenomena and for the use in nanoelectronics. The ground state of this heterostructure with the AA stacking of graphene layers is found to correspond to the maximum in the dependence of the tunneling conductance between the graphene layers on their in-plane relative displacement. Therefore, any relative displacement of the layers causes a decrease of the tunneling conductance and, thus, this heterostructure can be perspective for elaboration of nanoresonator-like sensors based on measurements of the amplitude of relative in-plane vibrations of the graphene layers through measurements of the tunneling conductance between them (such sensors were proposed on the basis of double-walled carbon nanotubes[71,82]). This is different from the bilayer graphene with the AB stacking in the ground state where the direction of relative in-plane displacement of the layers determines whether a decrease or an increase of the tunneling conductance takes place.

Based on the potential energy relief describing relative in-plane displacements of the graphene layers with the commensurate krypton spacer calculated within the DFT-D2 method, the shear mode frequency for this heterostructure is estimated to be $f = 10.5 \, \text{cm}^{-1}$. Recently Raman measurements of the shear mode of few-layer graphene were reported.[72] The analogous measurements for the considered heterostructure could be used to test the calculations performed and, therefore, to test the adequacy of the van der Waals-corrected density functional theory for consideration of the interaction between graphene and inert gases.

**ACKNOWLEDGEMENTS**

AMP, AAK and YEL acknowledge support by the Samsung Global Research Outreach Program. AMP and AAK acknowledge support by the Russian Foundation for Basic Research (grant 11-02-00604). AMP and YEL acknowledge support by the Russian Foundation for Basic Research (grant 12-02-90041-Bel). YEL acknowledges support by a MIEM VShE Program. IVL acknowledges support by the Marie Curie International Incoming Fellowship within the 7th European Community Framework Programme (Grant Agreement PIIF-GA-2012-326435 RespSpatDisp), Grupos Consolidados del Gobierno Vasco (IT-578-13) and computational time on the Supercomputing Center of Lomonosov Moscow State University[83] and the Multipurpose Computing Complex NRC "Kurchatov Institute".[84] NAP, AIS and SAV acknowledge support by the Belarusian Republican Foundation for Fundamental Research (grant F12R-178) and Belarusian scientific program "Convergence".